\documentclass[12pt]{iopart}
\usepackage[dvips]{graphicx}
\usepackage{amssymb}

\begin{document}
\def\picbox#1#2{\fbox{\vbox to#2{\hbox to#1{}}}}
\def\bra#1{\langle#1|}
\def\ket#1{|#1\rangle}
\def\parc#1#2{\frac{\partial #1}{\partial #2}}
\def\rot{\textrm{rot}}
\def\grad{\textrm{grad}}
\def\pa#1{\partial_{#1}}
\def\scalar#1#2{\langle#1|#2\rangle}
\def\tr#1{\textrm{tr}\left\{#1\right\}}
\def\mean#1{\overline{#1}}
\def\ave#1{\left\langle #1\right\rangle}


\title{Wigner function statistics in classically chaotic systems}

\author{Martin Horvat and Toma\v z Prosen}

\address{Physics Department, Faculty of Mathematics and Physics, 
University of Ljubljana, Slovenia}

\eads{\mailto{martin@fiz.uni-lj.si}, \mailto{prosen@fiz.uni-lj.si}}  
\begin{abstract}
We have studied statistical properties of the values of the Wigner function $W(x)$ of 1D quantum maps on compact 2D phase space of finite area $V$. For this purpose we have defined a Wigner function probability distribution $P(w) = (1/V) \int \delta (w-W(x)) dx$, which has, by definition, fixed first and second moment. In particular, we concentrate on relaxation of time evolving quantum state in terms of $W(x)$, starting from a coherent state. We have shown that for a classically chaotic quantum counterpart the distribution $P(w)$ in the semi-classical limit becomes a Gaussian distribution that is fully determined by the first two moments. Numerical simulations have been performed for the quantum sawtooth map and the quantized kicked top. In a quantum system with Hilbert space dimension $N(\sim 1/\hbar)$ the transition of $P(w)$ to a Gaussian distribution was observed at times $t\propto \log N$. In addition, it has been shown that the statistics of Wigner functions of propagator eigenstates is Gaussian as well in the classically fully chaotic regime. We have also studied the structure of the nodal cells of the Wigner function, in particular the distribution of intersection points between the zero manifold and arbitrary straight lines.
\end{abstract}
\submitto{\JPA}
%

%
\section{Introduction}

The Wigner function (WF) \cite{wigner} is an essential concept of phase space representation of quantum mechanics, namely it is a useful and faithful representation of a pure or mixed quantum state in terms of functions of canonical classical phase space variables. It has many applications in various branches of physics, in particular in quantum optics. However, WF cannot be interpreted as quantum phase space distribution as it can develop negative values, in particular due to well known oscillatory interference fringes following e.g. coherent wave-packet superpositions.

WF also played an important role in the realm of quantum chaology \cite{berry2,voros}. In particular, it has been conjectured that WF of stationary eigenstates of bounded dynamical systems in quasi-classical regimes localizes onto classically invariant components of phase space. For example, for classically regular phase space regions, like e.g. KAM tori, WF is supposed to become a Dirac delta function on a KAM torus whereas for a classically chaotic component WF is supposed to condense uniformly there. However, it is well known that this asymptotic behavior is to be understood in a very weak limit sense, namely WF is becoming 
uniform only after being integrated (with a very smooth test function) over non-small 
(or classical) region of phase space. On the other hand, on a smaller (quantum) scale, 
e.g. of Planck cell size, the phase space structure of WF of typical, 
say random or ergodic states, is very much unknown. This question is important for the 
understanding of decoherence and quantum stability with respect to system's perturbations 
as discussed recently \cite{zurek,PZ02,zurek2,Ben}.

In this paper we address the question of the structure of typical WF from a statistical point of view. We define and analyze the statistical distribution of values of WF of a given quantum state. In particular, we are interested in the relaxation of this distribution with time, when we start from initial coherent state, and in the corresponding time scale. Here we limit ourselves to classically fully chaotic and discrete time systems, namely the 
chaotic quantum maps, where the full phase space is the only topologically transitive ergodic component.

We show that for chaotic quantum maps, the limiting WF value distribution is a Gaussian. The average value is fixed by normalization, while the second moment (or the variance) diverges in the quasi-classical limit ($\hbar\to 0$), so we have roughly a symmetric distribution of positive and negative values of WF for random states. In addition, we show that for chaotic systems, the relaxation to equilibrium in statistics of WF happens on a short log $\hbar$ (Ehrenfest) time scale. Furthermore we show for chaotic systems, that the statistics of WF of typical chaotic eigenstates is the same as WF statistics of a random state, which is consistent with the established Berry's conjecture \cite{berry1}. Finally, we present statistical analysis of the structure of nodal cells of WF of chaotic or random states and show that it can be described by simplified models based on random trigonometric functions.

In section 2 we outline the essentials of general Wigner-Weyl formalism
and discuss two special cases of compact phase space, namely of toroidal and
spherical phase space, which are later used in numerical examples.
In section 3 we discuss WF value statistics of a random (or
quantum chaotic) state.
In secton 4 we discuss WF statistics on two numerical examples
of quantum chaos, and analyze in particular the relaxation of WF statistics in quantum
time evolution starting from 
an initial coherent state.
In section 5 we go beyond simple value statistics, and study also the
phase space structure statistics of WF, such as its
spatial correlation function and distribution of diameters of its nodal
cells, for example.
In section 6 we summarize our most important results and conclude.
\section{Weyl-Wigner formalism and the Wigner function}
Let us first review some of the essential general principles of Weyl-Wigner (WW) phase-space representation of operators (see e.g. \cite{deGroot}). As we are aiming at compact (finite) phase spaces of chaotic maps, our discussion of WW formalism has to be a little bit abstract such as to be able to incorporate different topologies that are discussed later. We begin by considering a phase space $\chi$ and a Hilbert space $\cal H$, such that a typical function $A(x), x\in \chi$, corresponds under quantization to a linear operator ${\hat A}$ over $\cal H$. The inverse map ${\hat A}\to A(x)$ is called a Weyl symbol, and if ${\hat A} = \ket{\psi}\bra{\psi}$ then the corresponding symbol is called the Wigner function (WF). The Weyl symbol can be formally constructed using a self-adjoint kernel operator ${\hat\omega}(x)$ with the property
\begin{equation}\label{eq:kernel1}
  \textrm{tr} \{\hat \omega (x)\hat \omega (y) \} = 
  \delta (x-y),\qquad
  \hat \omega(x)^\dag = \hat \omega(x),
\end{equation}
namely
\begin{equation}
  A(x) = \textrm{tr} \{ \hat\omega(x) \hat A \},
\label{eq:WWmap}
\end{equation}
which generates also the inverse map
\begin{equation}
  \hat A = \int dV A(x) \hat\omega(x).
\label{eq:WWimap}
\end{equation}
For the systems with classically compact (finite) phase space $\chi$ the corresponding Hilbert space is finitely, say $N$ dimensional. As a consequence, WW map (\ref{eq:WWmap}) can be uniquely inverted only if the phase space is restricted to a finite set of $N^2$ points $\chi'=\{x_{nk};n,k=0,1\ldots N-1\}$. Kernel is in this case defined in discrete points only, ${\hat\omega}_{nk} = {\hat\omega}(x_{nk})$ with the property
\begin{equation}\label{eq:kernel2}
  \textrm{tr} \{\hat \omega_{nk} \hat \omega_{ml} \} = 
  \delta_{nm}\delta_{kl},
\end{equation}
so that discrete WW map reads
\begin{equation}
  A_{nk} = \textrm{tr} \{ \hat\omega(x_{nk}) \hat A \},\quad\textrm{and}\quad
  \hat A = \sum_{nk} A_n \hat \omega_{nk}.
\end{equation}
The WF $W(x)$ is defined as a phase space representation of the density operator $\hat \rho$ multiplied by a certain suitable normalization constant $C$
\begin{equation}\label{eq:WFdef}
  W(x) = C \textrm{tr} \{ \hat \omega(x) \hat\rho \}.
\end{equation}
The constant $C$ can be set for the convenience of a particular application,
for example, usually it is set by the normalization of probability, $\int dx W(x) = 1$. 
However, in this paper we are interested in the fluctuation of the WF so we determine the
constant $C$ by fixing the standard deviation $\sigma^2 = \mean{W^2}-\mean{W}^2 = 1$, where $\mean{(...)}$ denotes the average over phase space. We note that the kernel $\hat{\omega}$ is far from being completely specified by the property (\ref{eq:kernel1}). In addition we need certain correspondence principles. For example, for the usual (non-compact) symplectic geometry $x=(q,p)$ in $d+d$ dimensional phase space the kernel reads  ${\hat\omega}(q,p) = (2\pi\hbar)^{-d/2} \int dv e^{{\rm i}p\cdot v/\hbar} \ket{q+v/2}\bra{q-v/2}$. However, for compact geometries the WW map may not be unique. In such cases one may decide for the most reasonable or the simplest choice. In any case, as effective value of Planck constant vanishes any consistent choice should yield equivalent results in the semiclassical limit.
\subsection{Wigner function on 2-d torus}
Many simple phenomena in the theory of bounded Hamiltonian (symplectic) dynamical systems can be demonstrated on a 2-d torus $\chi=T^2 = [0,2\pi) \times [0,2\pi)$. However, quantization of a torus is not trivial from mathematical point of view. Hence there are several different proposals for a WW formalism on a torus \cite{torfor}. Here we are following the reference \cite{agam} up to convenient scaling factors. Let the Hilbert space ${\cal H}$ have dimension $N$, and the canonical bases, namely the position basis $\ket{n},n = 0,\ldots,N-1$, and the momentum basis $\ket{\tilde k},k=0,\ldots,N-1$, satisfying $\scalar{n}{\tilde k} = N^{-1/2}\exp(-2\pi{\rm i}n k/N)$. 
The {\em quantum phase space} in our formalism is a discrete mesh of $N\times N$ points, $x_{nk} = (2\pi n/N,2\pi k/N)$. The WF function on such discrete phase space in the limit of $N\to\infty$ mimics the continuous Wigner function \cite{wigner}. Due to technical reasons we assume that $N$ is {\em odd}. The WW kernel in this formalism is defined as
$$
\hat \omega_{nk} = \frac{1}{\sqrt{N}}\sum_{n' l} 
\exp (-2\pi{\rm i} n' k /N) \tilde \delta (2l-2n+n') \ket{l}\bra{l+n'}
$$
where all indices run from  $-(N-1)/2$ to $(N-1)/2$, which will be assumed whenever we refer to the quantized torus. For a pure quantum state ${\psi}$ WF is given by
\begin{eqnarray}
  W_\psi(n,k) &=& C_t \tr{\hat\omega_{nk} \hat\rho}\qquad 
  C_t = \sqrt{\frac{N^3}{N-1}}\label{eq:WF_torus} \\
  &=& \frac{N}{\sqrt{N-1}}\sum_{n',l} 
    \exp [-2\pi{\rm i} n' k /N] \tilde \delta (2l-2n+n') 
    \scalar{l+n'}{\psi}\scalar{\psi}{l} \nonumber
\end{eqnarray}
where we define a fat Dirac delta function $\tilde \delta (l)$ as 
$$
  \tilde \delta (l) 
  = \frac{1}{N} \sum_{m'} \exp(\pi {\rm i} m' l/N) 
  = \frac{1}{N} \frac{\sin (\pi l /2)}{\sin (\pi l /2 N)}. 
$$

\subsection{Wigner function on a sphere}
As a second important special case we consider quantum mechanics of a spin $J$ variable whose classical phase space can be identified with a unit sphere $\chi=S^2$ described by spherical angles $\theta\in[0,\pi),\varphi\in[0,2\pi)$. The WW formalism for SU(2) geometry was developed in \cite{agarwal} with the kernel
$$
  \hat\omega(x) = \sum_{kq} \hat T_{kq} Y^{*}_{kq}(x),\qquad x=(\theta,\varphi),
$$
where $Y_{kq}$ are standard spherical harmonics and  $\hat T_{kq}$ are multipole operators defined by
$$
  \hat T_{kq} = \sum_{m=-J}^J 
           \sum_{m'=-J}^J (-1)^{J-m} 
	   \sqrt{2k+1} \pmatrix{J& k &J\cr -m& q & m'} \ket{Jm}\bra{Jm'}.
$$
Symbols \mbox{\tiny $\pmatrix{ J& k &J\cr -m& q & m'}$ } are standard Wigner 3j symbols \cite{angular}. Using the kernel one defines WF as \cite{downling} 
\begin{equation}\label{eq:WF_sphere}
  W(x)= C_s \tr{\hat \omega(x) \hat \rho} = 
  C_s \sum_{k=0}^{2J} \sum_{q=-k}^{q=k}G_{kq}Y_{kq}(x) , \quad 
  C_s = \sqrt{4\pi\frac{2J+1}{2J}}.
\end{equation}
The information about the quantum system state - density operator $\hat\rho$ is hidden in the coefficients
$$
  G_{kq} = \tr{\hat \rho \hat T_{kq}^\dag}.
$$
We note that SU(2) WF is defined continuously everywhere on a sphere.
However, it is a superposition of {\em finitely} many spherical functions with coefficients $G_{kq}$ which correspond to a discrete WF over a finite discrete mesh of points on some 
other compact geometries, like e.g. torus.

\section{Wigner function value statistics of a random state}

Quantum states of classically chaotic systems are usually associated to the so called {\em ergodic} or {\em random} wave functions. These random states can be constructed using very simple principles, like Berry's random plane wave superposition for billiards \cite{berry1}, or chaotic analytic functions in Bargman or Husimi representation of Hannay \cite{Hannay}, see also \cite{NV}. In this paper we want to analyze the Wigner function of a random state and its statistical properties. To best of our knowledge this has not yet been attempted before.

Starting from a fixed WF $W(x)$  we define its value distribution $P(w)$ as
\begin{equation}\label{eq:distrib1}
   P(w) = \frac{1}{V} \int_{\chi} \delta (w-W(x)) dx.
\end{equation}
We assumed that phase space $\chi$ is compact and has finite volume $V$. The averages with respect to the probability density $P(w)$ will be denoted by $\mean{(\ldots)}$. The first moment of $P(w)$ is fixed by normalization of Wigner function (given by constant $C'$). The second moment is also fixed by purity of the state which implies
\begin{equation}
  \int dx W^2(x) = C' \int dx W(x).
\end{equation}
Since in this paper we are interested in fluctuation of WF it is convenient to determine the scaling constant $C'$ (\ref{eq:WFdef}) by setting the standard deviation to one, $\sigma^2 = \mean{W^2} - \mean{W}^2 = 1$. This implies for the average values (first moments), using eqs. (\ref{eq:WF_torus},\ref{eq:WF_sphere}):
\begin{equation}
 \textrm{torus:} \quad \mean{W} = \frac{1}{\sqrt{N-1}},
 \qquad
 \textrm{sphere:} \quad \mean{W}= \frac{1}{\sqrt{2j}}.
\label{eq:mom1}
\end{equation}
It may be instructive to define a {\em relative standard deviation} in units of average WF $\kappa=\sigma/\mean{W}$. Then the above statement (\ref{eq:mom1}) says that the relative standard deviation $\kappa\to\infty$ of WF distribution {\em diverges} in the semi-classical limit $N\to\infty$ (or $j\to\infty$). 

We should note that the result on divergence of relative standard deviation is
general and independent of the structure of the state. So, if we have a very non-uniform state, e.g. Gaussian wave-packet, then the statement says merely that the standard deviation is very large compared to the average value since we have almost all density concentrated around a small region. However, if we have a random (ergodic) state, then the WF value distribution should be transitionally invariant in phase space. Therefore, the values of first two moments of global phase space WF value distribution should also determine the average and standard deviation of local distribution around each phase space point. In order to determine the entire WF value distribution we define an appropriate random WF model in a specific generic geometry, namely on a 2-d torus, although the result which will be obtained can be 
argued to be universal and geometry independent. We start with a random state in $N$ 
dimensional Hilbert space
\begin{equation}
  \ket{\psi} = \sum_{l=1}^N c_l \ket{l},\qquad  \sum_{l=1}^N |c_l|^2 =1
\label{eq:randomstate}
\end{equation}
where $c_l$ are random generally complex coefficients. It turns out (see e.g. \cite{haake}) that for large $N$, coefficient $c_l$ may be considered as independent complex Gaussian variables with variance $\ave{|c_l|^2}=1/N$ (here and later $\ave{\ldots}$ denotes averaging over an ensemble of random states (\ref{eq:randomstate})).

Translational invariance of WF value distribution implies that the distribution over phase space can be replaced with the distribution over ensemble of random states (\ref{eq:randomstate}), so we choose to study distribution of WF at the most easily computable phase space point $(0,0)$
\begin{equation}\label{eq:Wpoint}
  W_{00} = \frac{N}{\sqrt{N-1}}
    \left( 
      |c_0|^2 + \sum_{l\neq l'} \tilde\delta(l+l') c^*_l c_{l'}
    \right).
\end{equation}
In the leading order $\tilde\delta(l+l')$ can be approximated by Kronecker $\delta_{l+l'}$ so expression (\ref{eq:Wpoint}) is a sum of $N$ independent terms, the first $|c_0|^2$ is strictly non-negative, while the other have vanishing mean and finite fluctuation. Hence, due to central limit theorem, for large $N\gg 1$, the distribution of $W_{00}$ becomes Gaussian
\begin{equation}
P(w) = \frac{1}{\sqrt{2\pi}}\exp\left(-\frac{1}{2}(w-\mean{W})^2\right),
\label{eq:Gaussian}
\end{equation}
with the first and second moment which can be computed directly by ensemble averaging giving identical results to previous phase averaging (\ref{eq:mom1}), and $\sigma^2=1$.

\section{Dynamics and relaxation of Wigner function value statistics}
In this section we will consider dynamics, namely statistics of WF of time evolving pure states. In particular we shall focus on systems with ergodic, mixing, and fully chaotic classical dynamics such that in the course of classical dynamics any non-singular initial classical measure relaxes to a uniform (micro-canonical) measure. When turning to quantum mechanics we pose a simple problem. Let us start with 'the most classical' initial state, namely with the coherent state (e.g. Gaussian wave packet): First, how does WF value distribution of a time evolving state relax into the Gaussian distribution (\ref{eq:Gaussian}) which characterizes the final ergodic (random) state? Second, what is the characteristic time scale of this relaxation process for a typical chaotic system and how does it scale with the value of an effective Planck constant?

\subsection{Toy models of quantum chaotic dynamics}
In order to carry out the plan outlined above we need to define generic time evolution on the Hilbert spaces of the quantized torus or the quantized sphere, respectively, which are chaotic in the classical limit. This is not difficult as we simply consider popular models which have been widely studied in the literature, namely the quantized sawtooth map on the torus \cite{casati}, and the quantized kicked top \cite{haake}.

The quantized sawtooth map on the torus (classical counterpart is defined on a torus $(q,p)\in [0,2\pi]\times [0,2\pi L]$) defined on the finite Hilbert space of dimension $N$ with the evolution operator
\begin{equation}
  \hat U_s = \exp\left(-{\rm i}\frac{T}{2} {\hat m}^2\right) 
             \exp\left({\rm i}\frac{K_0 T}{2 L^2} {\hat n}^2\right)
\label{eq:sawtoothdef}
\end{equation}
where $K_0$ is a kicking strength, $T = 2\pi L/N$ is a period of forcing and integer $L$ (usually set to 1) measures the vertical size of toroidal phase space. We have introduced formal position and momentum operators with integer eigenvalues, namely $\hat n$ and $\hat m$ respectively, satisfying $\hat n \ket{n}= n\ket{n}$, $\hat m \ket{m}= m\ket{m}$. We should note that the classical sawtooth map is ergodic and uniformly hyperbolic for $K_0>0$. As for the initial state we always choose a coherent state (which can be expressed in terms of Jacobi theta functions for toroidal phase space \cite{torcohst}) centered somewhere on the torus.\par
As for the other system defined (classically) on a spherical phase space, the quantum kicked top of spin $J$ has a unitary map $\hat U_k$ acting on a $(2J+1)$-dimensional Hilbert space
$$
  \hat U_k = \exp\left(-{\rm i}\gamma \hat J_x\right)
             \exp\left({\rm i}\alpha \hat J_z^2/2J\right).
$$
Here, parameter $\alpha$ is an angle of rotation in between the kicks and $\gamma$ is the kicking strength. In the paper we will consider the parameter values $\gamma=\pi/2$ and $\alpha=3$, or $\alpha=10$, corresponding to classically mixed, or fully chaotic, phase space, respectively. Again, we prepare the system initially in the SU(2) coherent state \cite{SU2cohst} placed somewhere on the
sphere.
\par
\begin{figure}[!htb]
  \begin{center}
    \includegraphics[angle=-90, width=12cm]{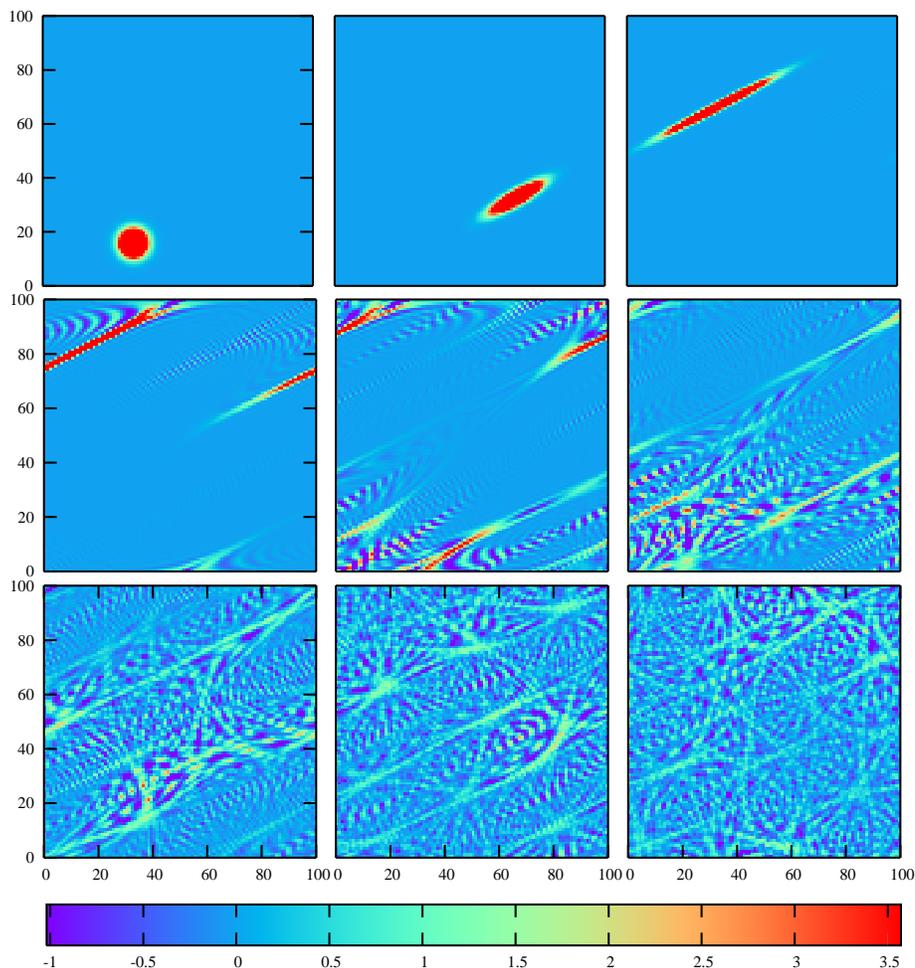}
  \end{center}
  \caption{The dynamics of WF in the quantized saw-tooth map for parameters 
    $K = 0.5$, $L = 1$ and dimension $N = 101$. The system is initially in the coherent state centered at $(q,p)=(2/3\pi,1/3\pi)$. WF at successive integer time steps is shown from left to right and from top to bottom. The color code bar of the Wigner function values is shown at the bottom of the figure. Note the significant contribution of negative values for longer times.}
\label{pic:saw_din}
\end{figure}
For the purpose of illustration we first show the cascade of snapshots of WF starting from initial coherent state undergoing quantum dynamics with completely chaotic classical time evolution. The results are shown in figure \ref{pic:saw_din} for the quantized sawtooth map while similar-looking cartoons were obtained for kicked top. It is clear from the figure that after very short time (of the order of few kicks) the Wigner function relaxes into a universally looking distribution with roughly symmetric distribution of positive and negative values. We conjecture that this asymptotic distribution should be a Gaussian (\ref{eq:Gaussian}) as derived for a random model. Furthermore, we conjecture that the time-scale on which this relaxation takes place is just of the order of the so-called Ehrenfest time \cite{BermanZaslavsky}, that is the time needed that the initially localized wave-packet spreads over the accessible phase space $t_{\rm ehr} = \log\hbar/\lambda$ where $\lambda$ is a classical Lyapunov exponent.

\subsection{Wigner function value distribution}

\begin{figure}[!htb]
  \begin{center}
    \includegraphics[angle=-90, width=12cm]{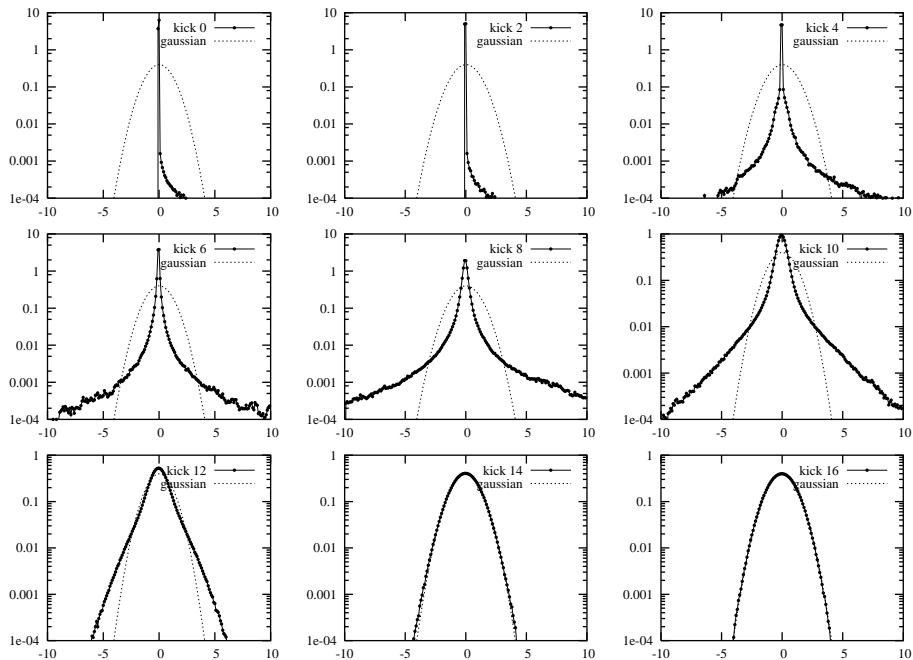}
  \end{center}
  \caption{Time evolution of the Wigner function value distribution (\ref{eq:distrib1}) of the quantized sawtooth 
map for $K_0=0.5$, $L=1$ and $N=2187$. Initial state is a coherent wave packet. Dashed curves give a theoretical Gaussian (\ref{eq:Gaussian}).}
\label{pic:saw_dist}
\end{figure}

\begin{figure}[!htb]
  \begin{center}
    \includegraphics[angle=-90, width=14cm]{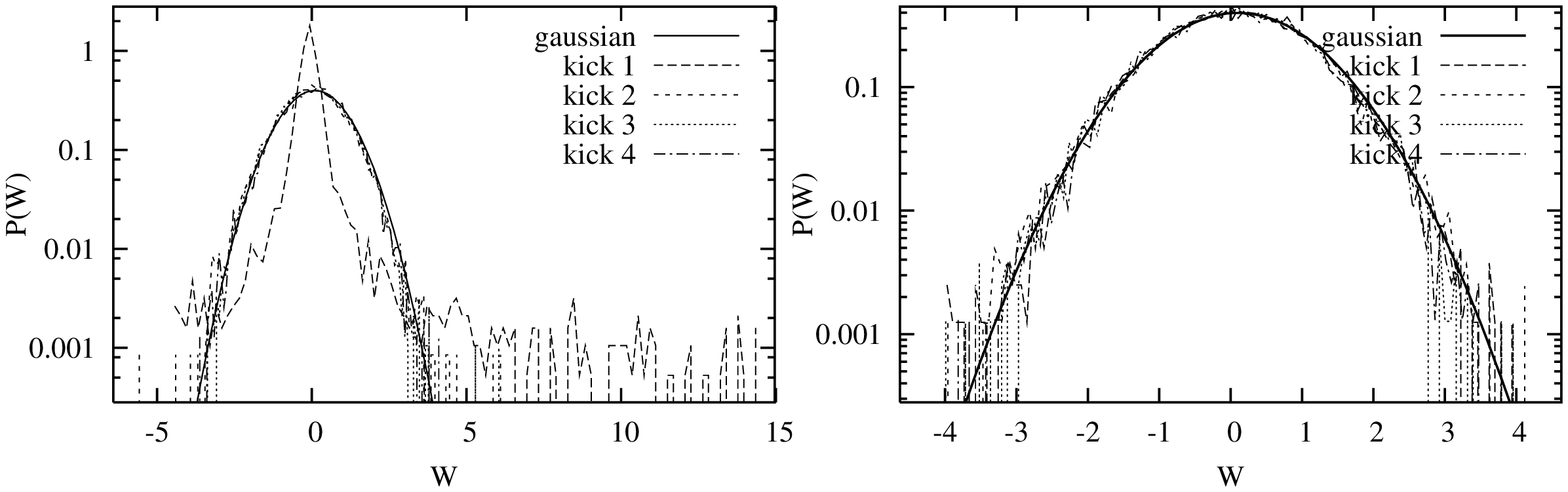}
  \end{center}
   \hfil (a)\hfil\hfil (b) \hfil
  \caption{Time evolution of the Wigner function value distribution (\ref{eq:distrib1}) of the quantized chaotic kicked top for $\alpha=10$, $\gamma=\pi/2$ and $J=50$. In (a) we take initial coherent state whereas in (b) we take initial random state. Consecutive kicks are show with different line styles and the full curve is a theoretical Gaussian (\ref{eq:Gaussian})).}
\label{pic:kt_dist}
\end{figure}

Firstly, we want to check the relaxation of WF value statistics $P(w)$ starting from the initial coherent state. Using numerical experiments with the sawtooth map and the kicked top we confirm our expectation and obtain fast relaxation into statistically significant Gaussian distribution (\ref{eq:Gaussian}). The cascade of distributions $P(w)$ for several consecutive kicks in the sawtooth map is shown in figure \ref{pic:saw_dist}. It is perhaps interesting to note that the relaxation goes through an intermediate distribution which seems to have exponential tails (see a snapshot at 10th kick). A very similar result is obtained also for the kicked top model shown in figure \ref{pic:kt_dist}a. For comparison we show in figure \ref{pic:kt_dist}b stationary WF value distributions obtained from time evolution starting from a random state. In this case we obtain, as expected, for the initial state, and for time evolving states, a nice agreement with a Gaussian WF value distribution.

\subsection{Relaxation time scale} \label{sec:time_scales}

Secondly, we want to quantitatively characterize the deviations from a Gaussian statistics and thus to measure the time scale of relaxation. To this end we define the excess $\epsilon$ of WF value distribution $P(w)$
\begin{equation}
 \epsilon = \mean{(w-\mean{W})^4}/\sigma^4-3.
\label{eq:excess}
\end{equation}
Note, that $\epsilon = 0$ for a Gaussian (\ref{eq:Gaussian}) and the size of $\epsilon$ should roughly measure deviation from a Gaussian. As we have discussed above, we expect the relaxation process to take place within the Ehrenfest time scale
\begin{equation}
  t_{\rm r} \sim \log N/\lambda,
\label{eq:trelax}
\end{equation}
where $\lambda$ is an effective (average) Lyapunov exponent and $N$ is a dimensionality of Hilbert space. This is certainly a lower bound to a relaxation time scale, but as we will show by numerical experiments, it also gives the right scaling and order of magnitude of the true relaxation time. For this purpose we choose only the quantized sawtooth map where much larger $N$ is accessible so that the logarithmic scaling can be checked.\par
\begin{figure}[!htb]
  \begin{center}
    \includegraphics[angle=-90, width=10cm]{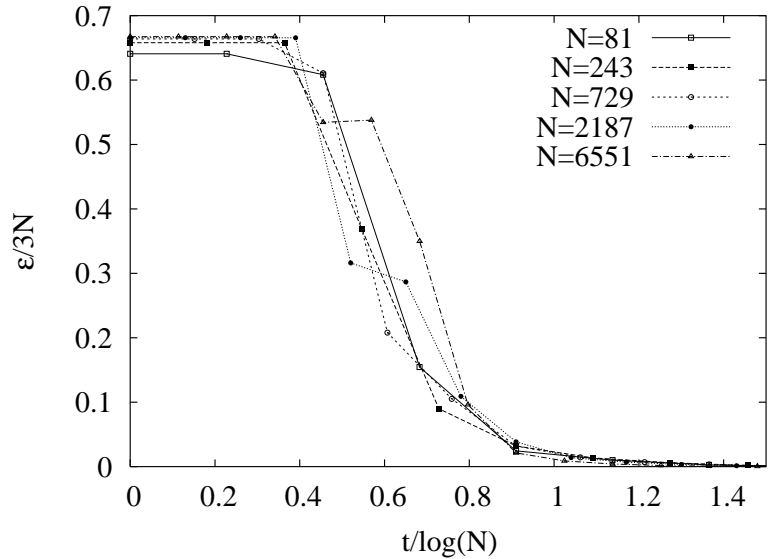}
  \end{center}
\caption{Scaling of relaxation to Gaussian statistics in the quantized saw-tooth map, $K_0=0.5$, $L=1$, measured with the excess $\epsilon$ of the 
distribution. The initial state is a coherent wave packet. 
Different dimensions $N$ are indicated in the figure.}
\label{pic:saw_mom}
\end{figure}
\begin{figure}[!htb]
  \begin{center}
    \includegraphics[angle=-90, width=9cm]{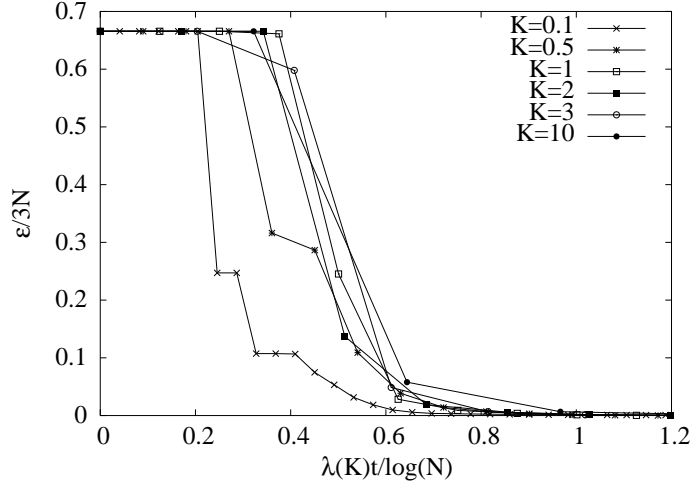}
  \end{center}
  \caption{Excess $\epsilon$ as a function of time for the quantized saw-tooth map at different kicking strength $K_0$ (see figure) and at fixed dimension $N=2187$ and at $L=1$. Lyapunov exponent is given by the formula: $\lambda(K_0)=\log((2+K_0+((2+K_0)^2-4)^{1/2})/2)$ 
\cite{casati}.}
\label{pic:saw_cmp}
\end{figure}

In figure \ref{pic:saw_mom} we plot $\epsilon$ as a function of time of WF value distribution starting from a coherent initial states, for different values of $N$ over several orders of magnitude. Indeed we observe very clean transition from $\epsilon \approx 0.65$ to $\epsilon = 0$ at around $t\approx\log N$. In order to check also the dependence on the Lyapunov exponent $\lambda(K_0)$ we have in figure \ref{pic:saw_cmp} plotted the relaxation of excess $\epsilon$ for several different values of chaoticity parameter $K_0$ and fixed $N$. We have found that the transition time scale is proportional to $\lambda^{-1}$ therefore supporting the formula (\ref{eq:trelax}).\par
\begin{figure}[!htb]
  \begin{center}
    \includegraphics[angle=-90, width=9cm]{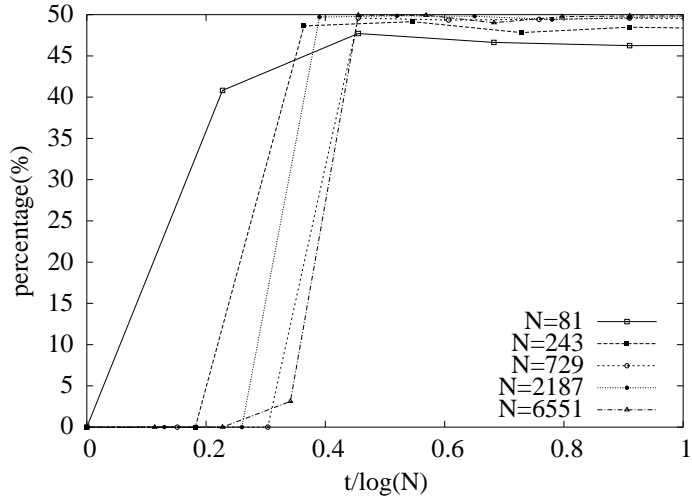}
  \end{center}
  \caption{Percentage of phase space with negative WF as a function of scaled
time for the quantized saw-tooth map, $K_0=0.5$, $L=1$ at several different dimensions $N$ as indicated in the figure.}
\label{pic:saw_perc}
\end{figure}

Since $\epsilon$ is only one number which cannot characterize the overall distribution $P(w)$ we have made the following additional check. We know that for a random state in the limit $N\to\infty$ the probability of having negative value of WF goes to $1/2$. Therefore we propose to study the percentage of phase space with negative valued WF 
\begin{equation}
  P_- =\int_{-\infty}^0 P(w) dw
\end{equation}
as a function of time. The results of numerical experiment for the quantized sawtooth map are shown in figure \ref{pic:saw_perc}. We see that $P_-$ converges very fast to $1/2$ on a time scale ($t_{\rm c}$) proportional to $\log(N)$ but significantly (around a factor of two) smaller than the relaxation time $t_{\rm r}$. We note that negative values of WF mean serious deviation from classical Liouville density, so our result indicates that quantum classical correspondence breaks significantly before the WF value distribution becomes actually close to a Gaussian. However, both time scales have the same scaling with $N$ and
$\lambda$.
\subsection{Value statistics of Wigner functions of stationary eigenstates}

\begin{figure}[!htb]
  \begin{center}
    \includegraphics[angle=-90, width=9cm]{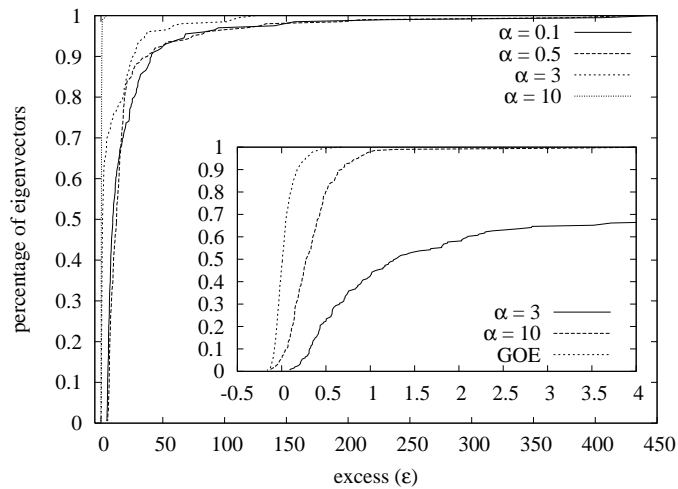}
  \end{center}
  \caption{Cumulative distribution of the excess $\epsilon$ of the quantum kicked top eigenfunctions in the case of classically (almost) regular ($\alpha=0.1, 0.5$), mixed ($\alpha=3$), and fully chaotic ($\alpha=10$) dynamics. To compare the classically chaotic case with random matrix theory, the result for GOE random matrix eigenfunctions is given in the inset in the blown up scale.}
\label{pic:kt_eigen}
\end{figure}

We note that the time evolution of a Wigner function of initial state
\begin{equation}
 \ket{\psi}= \sum_k a_k  \ket{\psi_k}, \qquad 
  a_k\in \mathbb{C},\quad \sum_k |a_k|^2 = 1,
\end{equation}
can be written as
\begin{eqnarray}
  W (x,t) &=& \sum_{kl} a_k a_l^{*} \exp\left(i t (\Omega_k -\Omega_l)\right)
           W_{kl} (x), \label{eq:W_eigen}\\
  W_{kl} (x) &=& C\tr { \hat\omega (x) \ket{\psi_k}\bra{\psi_l} },\nonumber
\end{eqnarray}
in terms of Wigner function basis $W_{kl}$ if $\ket{\psi_k}$ are eigenfunctions of the propagator $\hat U^t$ with eigenvalues $\exp(it\Omega_k)$ . Therefore, stationary (or time averaged) properties of a WF $W(x,t)$ are determined by the diagonal functions $W_{kk}$, which are the usual WF of the eigenstates of $U$. It is therefore clear that for classically chaotic and ergodic systems one should expect {\em almost all} $W_{kk}(x)$ to have the property of a WF of a random state. In order to confirm this conjecture we have computed the $\epsilon_k$ of WF value distribution of a complete set of eigenfunctions $\{W_{kk}(x),k=1,2,\ldots\}$ and analyzed its distribution. Indeed we found, as shown in figure \label{pic:ki_eigen} for the quantized kicked top, that in the classically chaotic case almost all eigenfunctions have $\epsilon\approx 0$. This has been further quantitatively compared to Random Matrix Theory by computing excess distribution for Wigner functions of eigenvectors of a Gaussian orthogonal (GOE) random matrix. Indeed very nice agreement was found, meaning that also the number of Wigner functions with larger excess ('accidental scars') is within the statistical fluctuation predicted by Random Matrix Theory. In the same figure (\ref{pic:kt_eigen}) we also analyze excess distribution for mixed and regular classical dynamics, where the excess distribution has a nontrivial shape.

\subsection{Auto-correlation of the Wigner function}\label{sec:corr}
\begin{figure}[!htb]
  \begin{center}
    \includegraphics[angle=-90, width=12cm]{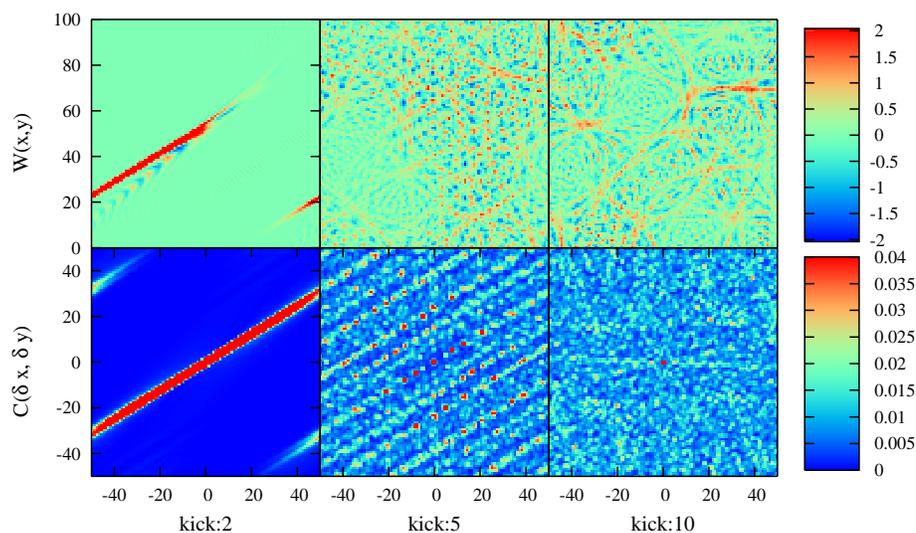}
  \end{center}
  \caption{The density plot of autocorrelation $C(\delta x,\delta y)$ (below) and Wigner function $W(x,y)$ (above) in time evolution of the sawtooth map at dimension $N=101$ and $K_0=1$, $L=1$.}
\label{pic:saw_corr}
\end{figure}

Statistical description of WF at individual points in phase space is meaningful if statistical correlations between Wigner function values at different phase space points are small. Therefore it may be interesting to define and study the auto-correlation function of Wigner function, defined as
\begin{eqnarray*}
  C(\delta x) &=& 
       \frac{1}{V}\int W(x) W(x+\delta x) dx 
  ~\textrm{(continuous phase space)},\\
  C(\delta n,\delta m) &=& 
       \frac{1}{N^2}\sum_{nm} W_{nm} W_{n+\delta n,m+\delta m}   
   ~\textrm{(discrete phase space)}.
\end{eqnarray*} 
The dynamics of the auto-correlation and the corresponding Wigner function starting from the coherent state placed at ($q=2.1,p=1.2$) of quantized saw-tooth are shown in figure \ref{pic:saw_corr}. At small times, when Wigner function has still a clear non-random structure, the auto-correlation varies strongly with the displacement vector $\delta x=(\delta m,\delta n)$. From autocorrelation function one can clearly observe the directions of stable and unstable classical flow. At later times $C(\delta x)$ becomes isotropic, and apart from the delta spike at the origin $\delta x=0$, equal a constant, namely $C(\delta x) \sim \delta(\delta x) + \textrm{const.}$. Indeed, for a random state we shall proceed to determine $C(\delta x)$ exactly.\par
\begin{figure}[!htb]
  \begin{center}
    \includegraphics[angle=-90, width=9cm]{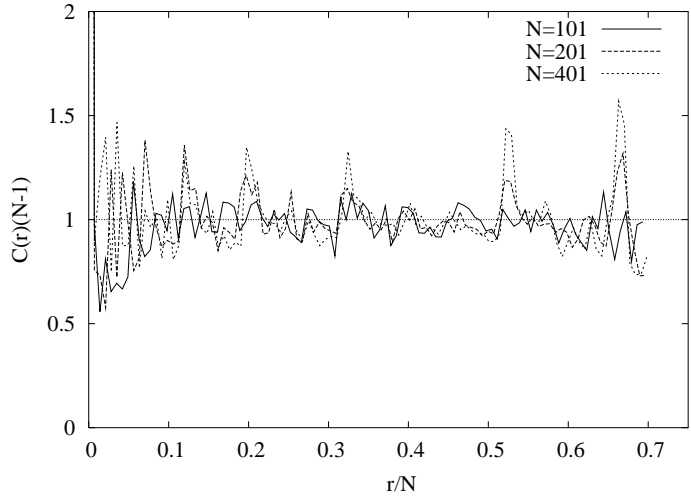}
  \end{center}
  \caption{Scaled local time average auto-correlation function $\mean{C}$ versus the radial distance $r=\sqrt{(\delta x)^2+(\delta y)^2}$ for different dimensions $N$ of the quantized sawtooth map for $K_0=1$ and $L=1$.}
\label{pic:saw_corr_model}
\end{figure}

We assume a random wave hypothesis, namely that a random (time-dependent) state can be written as $\ket{\psi}=\sum_n c_n\ket{n}$, where $c_n$ are Gaussian random independent complex coefficients, satisfying $\ave{c_n^* c_m} = \delta_{nm}$, $\ave{c_n c_m} = 0$, where $\ave{\ldots}$ represents the average over an ensemble of states, or equivalently, a time-average with a random initial state. The auto-correlation function can thus be expressed as
\begin{eqnarray}
   C(\delta n,\delta m) = \frac{N}{N-1} |\alpha(\delta n,\delta m)|^2 
   \label{eq:corr_psi}, \\
   \alpha(\delta n ,\delta m) = \sum_l \exp(-{\rm i} 2\pi l \delta m/N) 
   c_l c_{l+\delta n}^{*}. \nonumber
\end{eqnarray}
Averaging over Gaussian random variables $c_n$ can be performed straightforwardly, yielding
$$
  \ave{C(\delta n,\delta m)} 
  =\frac{N}{N-1}\delta_{\delta n,0} \delta_{\delta m,0} + \mean{W}^2.
$$
This theoretical prediction is in good agreement with numerical experiment, as shown in figure \ref{pic:saw_corr_model}. We also estimate (temporal) fluctuations of $C(\delta x\neq 0)$ by standard deviation $\sigma^2_C = \ave{C^2} - \ave{C}^2$, which can be computed directly using Wick pair contractions in the variables $c_n$
$$
  \sigma_C = \ave{C} = \frac{1}{N-1}= \mean{W}^2\qquad {\rm if} \qquad
  (\delta n,\delta m) \neq 0,
$$
in the asymptotic regime of high dimension $N\gg 1$. 
\section{Wigner function phase space structure statistics}
In previous sections we have been investigating the value statistics of the Wigner function. WF possesses hills and valleys as compared to the reference ``altitude'' --- offset. Here we discuss 2d compact phase space where these structures are supported by 2d nodal cells, that in general have very rich topology. The nodal cell sizes, especially the sub-Planck size structures, have recently been brought in the connection with decoherence \cite{zurek,srednicki}.
We are now interested in the statistical properties of WF phase space structures such as the nodal cell size and the amplitude of oscillations (hills and valley) from a given offset value. For the offset we choose the phase space mean value of WF (offset = $\mean{W}$) as this seems to be the most natural choice. However, since the relative mean value divided by the fluctuation goes to zero, $\mean{W}/\sigma_W \to 0$ in the limit $N\to\infty$, we argue that the results are asymptotically just the same as if one considers nodal cells with respect to zero offset.

In this work we would like to obtain some general results on WF nodal cell statistics of random (or chaotic) states. In our numerical analysis we consider time-evolving states, starting from initial coherent state, for times $t > t_E$, when the statistics is stationary and the structure of the state is the expected to be the same as that of a random state.\par
\begin{figure}[!htb]
  \begin{center}
    \includegraphics[angle=0, width=8cm]{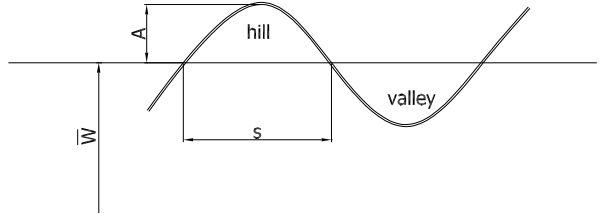}
  \end{center}
  \caption{The sketch of 1D phase space structures of the intersected WF.}
\label{pic:str_sketch}
\end{figure}
In order to simplify the picture we have considered the statistical properties of intersections between nodal cells (bounded by the curves $W(x)-\mean{W}=0$) and some arbitrary (random) straight lines in phase space, i.e. we consider one-dimensional projections of two-dimensional WF. Let the straight line $R(t)\in \chi$ be parametrized by a real variable $t$. In the case of spherical phase space, the set of all straight lines consists of all big circles, while in the case of toroidal phase space we consider for simplicity only closed straight lines, i.e. the two sets of irreducible circles specified by either fixing position or fixing momentum.

The central object studied here is the so-called Wigner function on a line (abbreviated WFL) with an offset value subtracted and assuming that the circumference of the circle is equal to $2\pi$:
\begin{equation}
  \widetilde W_{\psi,R} (t) = W_\psi [R(t)] - \mean{W},\qquad R(t) \in \chi, \quad t\in [0,2\pi].
\end{equation}
Since the statistical properties of a Wigner function of a random or ergodic quantum state are invariant under phase space translations, the parameters of the line (circle) $R(t)$ can also be chosen as random. In other words one may average over lines $R(t)$ in a uniform way, such that an ensemble of lines $R(t)$ uniformly covers the phase space with respect to an ergodic invariant measure. We shall investigate the following statistical distributions defined with respect to $W_{\psi,R} (t)$: (i) distribution $d{\cal P}(s)/ds$ of spacings $s=t_{n+1}-t_n$ between adjacent zeros $t_n$ of $W_{\psi,R}(t)$ which may also be called 'diameter distribution of Wigner nodal cells', (ii) distributions of amplitudes $d{\cal P}(A)/dA$, i.e. local maxima (hills) and local minima (valleys) between each pair of adjacent zeros as illustrated in figure \ref{pic:str_sketch}, and (iii) joint distribution of spacings $s$ and the corresponding amplitudes $A$, $d{\cal P}(s,A)/dsdA$.
\subsection{Random model of a Wigner function on a line}
We would like here to propose a simple statistical model which reproduces the properties of a random Wigner function on a line. We start by a simple ansatz expanding WFL into the Fourier modes 
\begin{equation}
  \widetilde W(t) = u_0 + \sum_{q=1}^M u_q \cos(q t) + v_q \sin(q t), \label{eq:model}
\end{equation}
where coefficients $u_q$ and $v_q$ are some Gaussian (asymptotically, as $N\to\infty$, statistically independent) random variables with zero mean $\ave{u_q}=0, \ave{v_q}=0$ and prescribed variances $\ave{u_q^2} = O(N^{-1}),\ave{v_q^2} = O(N^{-1})$. The effective number of Fourier modes $M$ is usually of the same order than the dimension of the Hilbert space $N$.
The ansatz (\ref{eq:model}) shall be proven separately for the case of WF on the torus and on the sphere, and also the expressions for the variances of coefficients $u_n$ and $v_n$ shall be computed. We will then use our statistical model to compare with exact numerical calculations in our two model systems.
\subsubsection{Random model on the torus}
\begin{figure}[!htb]
  \begin{center}
    \includegraphics[angle=-90, width=7cm]{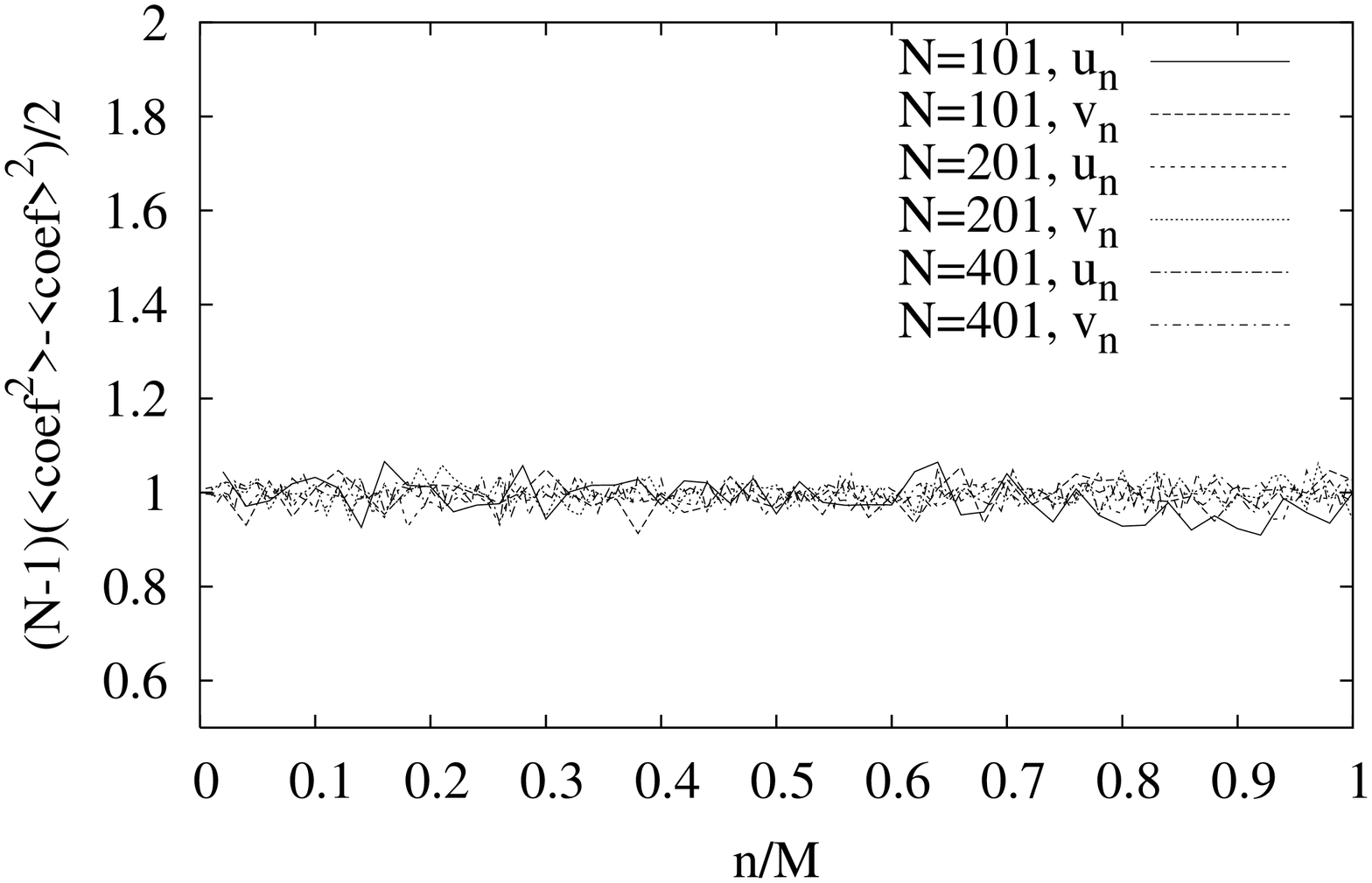}
    \includegraphics[angle=-90, width=7cm]{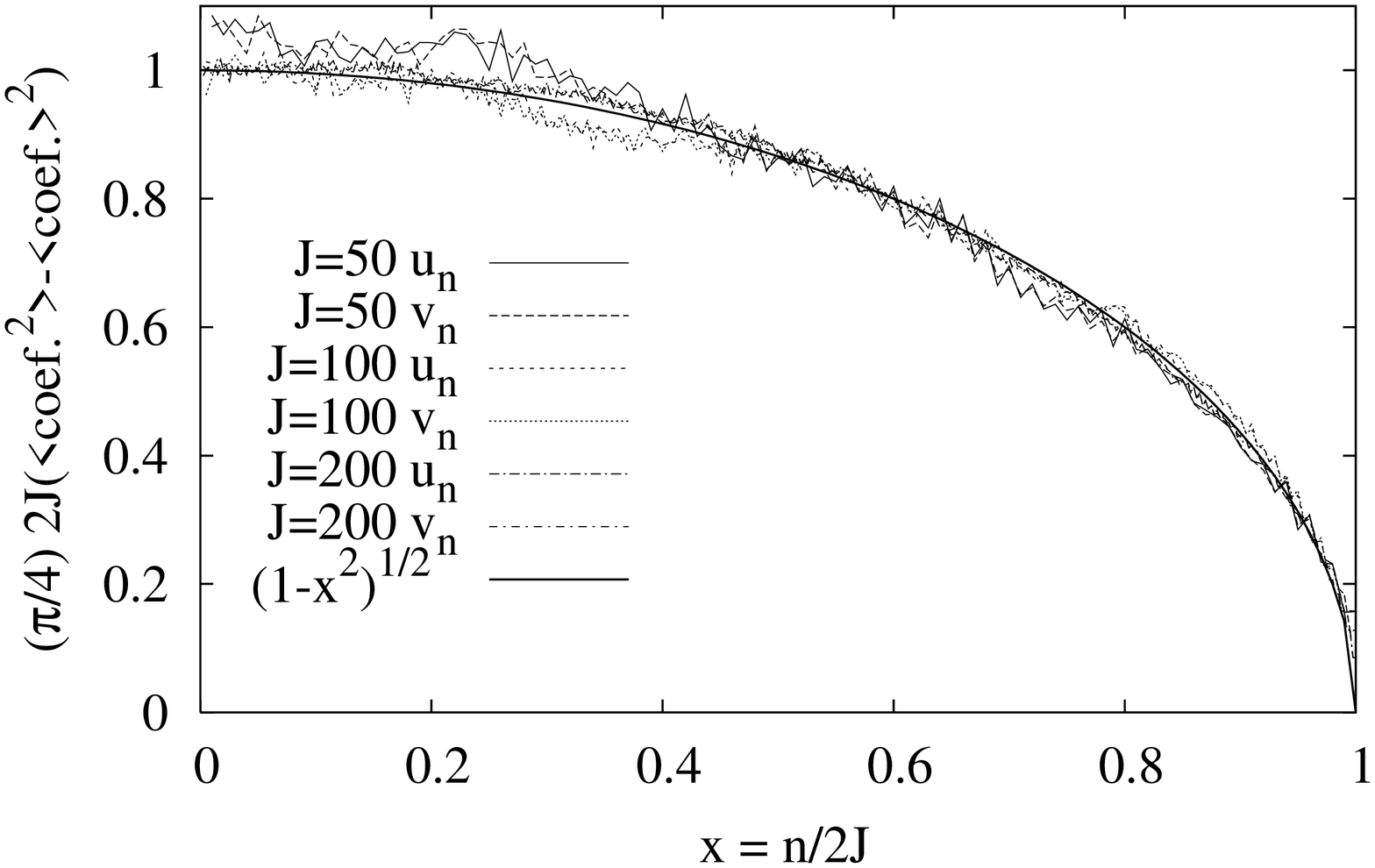}
  \end{center}
  \hfil (a) \hfil\hfil (b) \hfil 
  \caption{Variances of the Fourier coefficients of WFL of a random state: (a) in case of toroidal phase space, and (b) in case of spherical phase space.}
\label{pic:str_koef}
\end{figure}

We expect that statistical structure of random WF to be the same in both simplest sets of directions, namely for fixed position or fixed momentum. We limit the analysis which follows to the circles of fixed position. We start from a 'random' state written in the position basis 
$\{\ket{n}\}_{n=1,\ldots,N}$ of the Hilbert space of dimension $N$, as
\begin{equation}
  \ket{\psi} = \sum_{n=0}^N c_n \ket{n},\qquad 
  c_n \in \mathbb{C},
\end{equation}
with uncorrelated complex Gaussian coefficients (in the asymptotic regime $N\to\infty$) specified by
\begin{equation}
\ave{c_n}=0, \quad \ave{c_m^* c_n} = \frac{1}{N}, \quad \ave{c_m c_n} = 0.
\end{equation}
The expression of WF (\ref{eq:WF_torus}) at some fixed position $n$ can be rewritten in terms of {\em continuous} momentum variable $t = 2\pi m/N \in [0,2\pi)$ 
\begin{equation}\label{eq:model_torus}
 W_{\psi,n}(t) =\sum_{q=-M}^M Z_q(n) \exp({\rm i}q t),\quad M = (N-1)/2.
\end{equation}
Complex Fourier expansion coefficients $Z_q$ are expressed in terms of state coefficients $c_n$
\begin{equation}
 Z_q(n) = \frac{N}{\sqrt{N-1}} \sum_{l=-M}^M\tilde \delta (2l - 2n - q) c_{l-q} c_l^*,\quad Z^*_q(n) = Z_{-q}(n).
\label{eq:Zq}
\end{equation}
Now, the Fourier modes in (\ref{eq:model}) are a simple bilinear functions of the coefficients $c_n$,
namely
\begin{equation}
  u_0 = Z_0 - \mean{W},
  \quad
  u_q = Z_q + Z^*_q,
  \quad
  v_q = {\rm i}(Z_q - Z^*_q),
  \quad 
  q = 1,2\ldots M.
  \label{eq:Fcoef}
\end{equation}
By means of the central limit theorem one can argue in the asymptotic regime of large $N$, due to summation of statistically independent terms in eq. (\ref{eq:Zq}), that $u_n$ and $v_n$ become (independent) Gaussian variables of vanishing first moments $\ave{u_n}=0$, $\ave{v_n}=0$. Straightforward but tedious calculation gives for the second moments
\begin{equation}
  \ave{u_q u_{q'}} = \ave{v_q v_{q'}} = \frac{2}{N-1}\delta_{q,q'},\quad 
  \ave{v_q u_{q'}} =0, \quad \ave{u_0^2} = \frac{1}{N-1}.
\label{eq:pred}
\end{equation}
Therefore, the variances of Fourier coefficients are asymptotically, as $N\to \infty$ independent of the mode number $q$. This has been also verified by means of numerical simulation of WFL of random states, where $c_n$ have been generated using a suitable random number generator. The results are shown in figure \ref{pic:str_koef}a, where one sees excellent agreement with the prediction (\ref{eq:pred}).
\subsubsection{Random model on the sphere}
All big circles on the sphere are statistically equivalent with respect to WF (\ref{eq:WF_sphere}) of a random state. Thus we choose to consider the equator $\theta=\pi/2$ for simplicity. Using explicit expression for the spherical harmonics $Y_{kq}$
\begin{equation}
  Y_{kq} (x) = N_{kq} P_k^q(\cos(\theta))\exp(iq\phi),
\end{equation}
where $N_{kq}$ are normalization constants and $P_k^q$ are generalized Legendre polynomials, we have explicit expression for the WF on the equator
\begin{equation}
  W_\psi(\phi) = \sum_{q=-2J}^{2J} Z_q \exp(iq\phi),
\end{equation}
in terms of complex Fourier coefficients $Z_q$
\begin{eqnarray*}
  Z_q        & = & C_s \sum _{m,m'= -J} c_m c_{m'}^* K_{m,m'}^q, \\
  K_{m,m'}^q & = & \sum_{k=|q|}^{2J} N_{kq} P_k^q(0) 
                  \bra{J m'} T_{kq}^\dag \ket{J m} \in \mathbb{R}.
\end{eqnarray*}
The coefficients $Z_q$ are bilinear forms of wave coefficients $c_m$ with the Gramm matrix $K_{mm'}^q$. The latter is difficult to evaluate explicitly. Using the symmetry of coefficients $Z_q=Z_{-q}^*$, we rewrite WFL as
\begin{equation}
  W_\psi(\phi) - \mean{W} = u_0 +\sum_{q=1}^{2J} 
   u_q \cos(q\phi) + v_q \sin(q\phi),
\end{equation}
where real Fourier coefficients are again simply related to complex coefficients $Z_q$ by (\ref{eq:Fcoef}).
%
The WF random model of spin $J$ system has $M=2J$ modes. In the semi-classical regime (large $J$) we can again show by means of central limit theorem that $u_q$ and $v_q$ should be Gaussian distributed. Taking into account the detailed properties of terms in $Z_q$ we can find the following statistical properties of $u_q$, $v_q$ and $u_0$:
\begin{eqnarray*}
  \ave{u_q} = \ave{v_q} = \ave{u_0} = 0,\quad 
  \ave{u_q v_{q'}} = 0,\\
  \ave{u_q u_{q'}} = \ave{v_q v_{q'}} =\sigma_q^2 \delta_{q,q'},\quad 
  q = 1,\ldots,2J.
\end{eqnarray*}
The standard deviations of coefficients $\sigma_q$ are essential for further discussion, but they are quite difficult to compute analytically. We therefore obtain them by numerical study of WFL for a random state. The results of numerical simulation are shown on the figure \ref{pic:str_koef}.b. For large $J$, the results seem to be perfectly fitted by the {\em semicircle law}
\begin{equation}\label{eq:cycle_dev}
  \sigma^2_q = \frac{2}{J\pi} \sqrt{1-\left(\frac{q-1}{2J}\right)^2},\qquad 
   q = 1,\ldots, 2J,
\end{equation}
which is conjectured to be the correct semi-classical limit.
\subsection{Numerical study of the structure statistics of Wigner function}
Here we report on numerical simulations of the structure statistics of WFL of time-dependent states in two classically chaotic quantum systems, namely of the quantized sawtooth map at $K_0=10$ on the torus, and quantized kicked top at $\alpha=10,\gamma=\pi/2$ on the sphere. In addition, we simulate for comparison the corresponding statistics for the appropriate random model discussed in the previous paragraphs, since these seem to be impossible to be expressed analytically.\par
\begin{figure}[!htb]
  \begin{center}
     (a)\includegraphics[angle=-90, width=7cm]{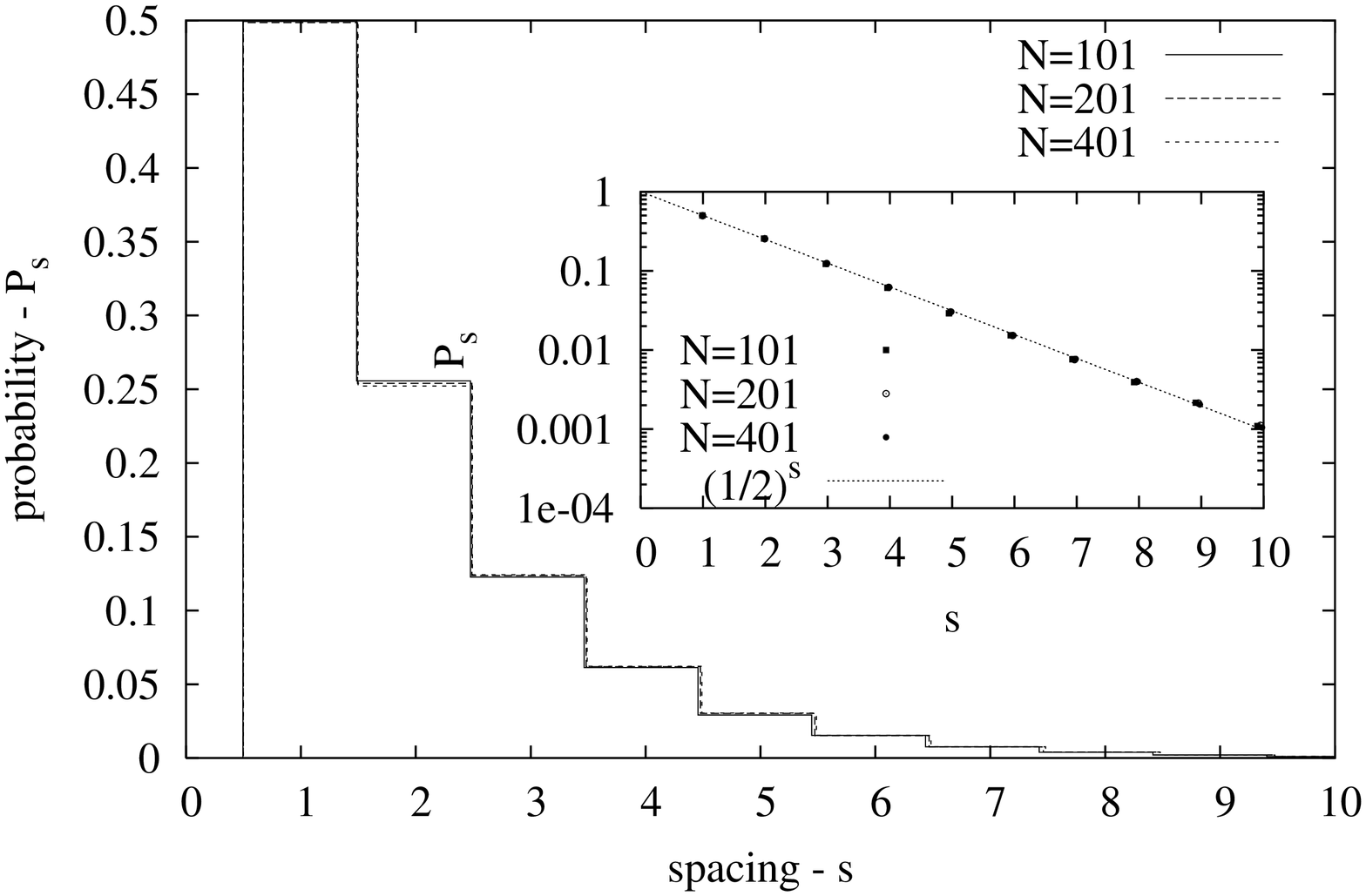}
     \includegraphics[angle=-90, width=7cm]{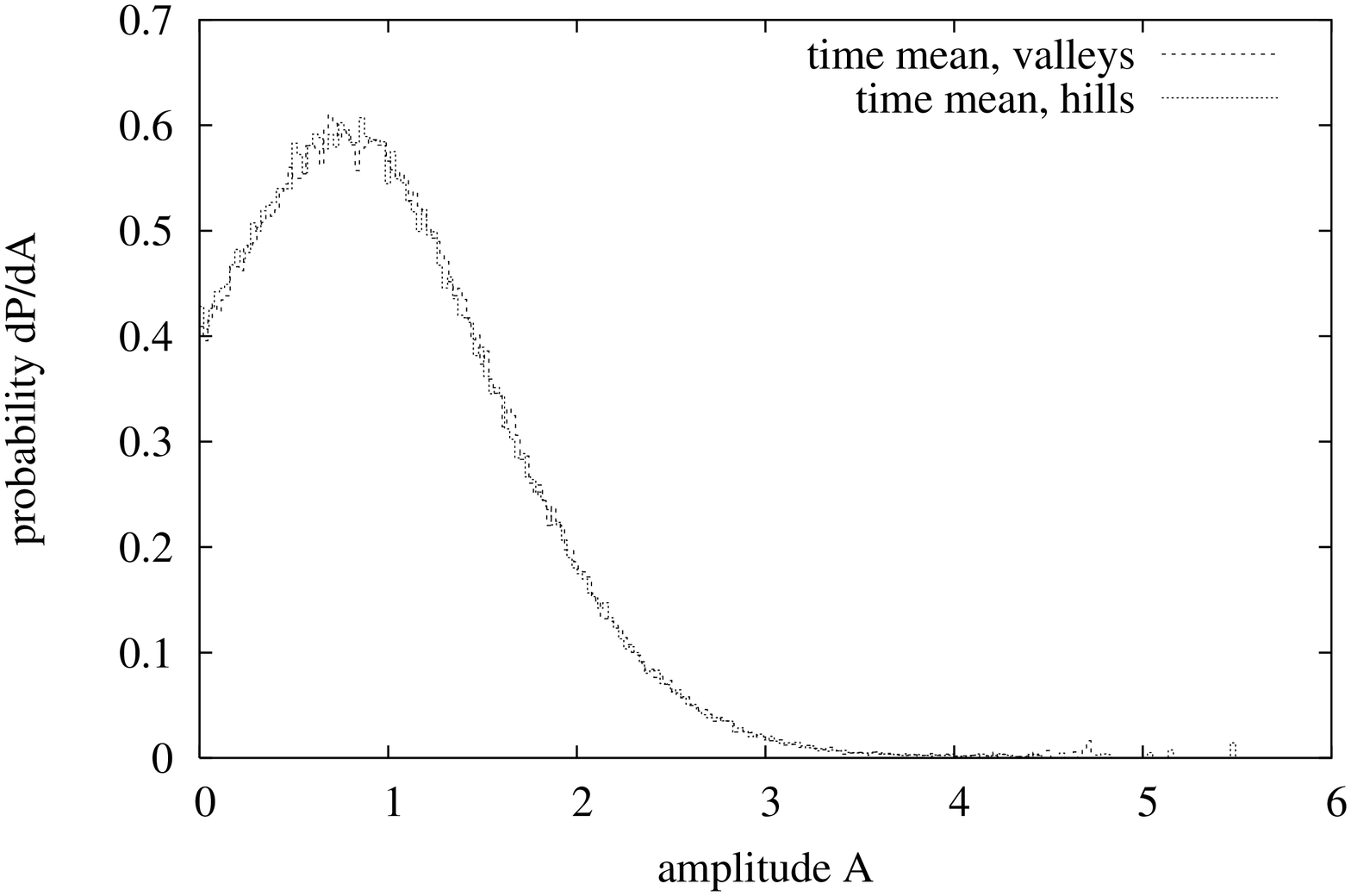}
    (b)\includegraphics[angle=-90, width=7cm]{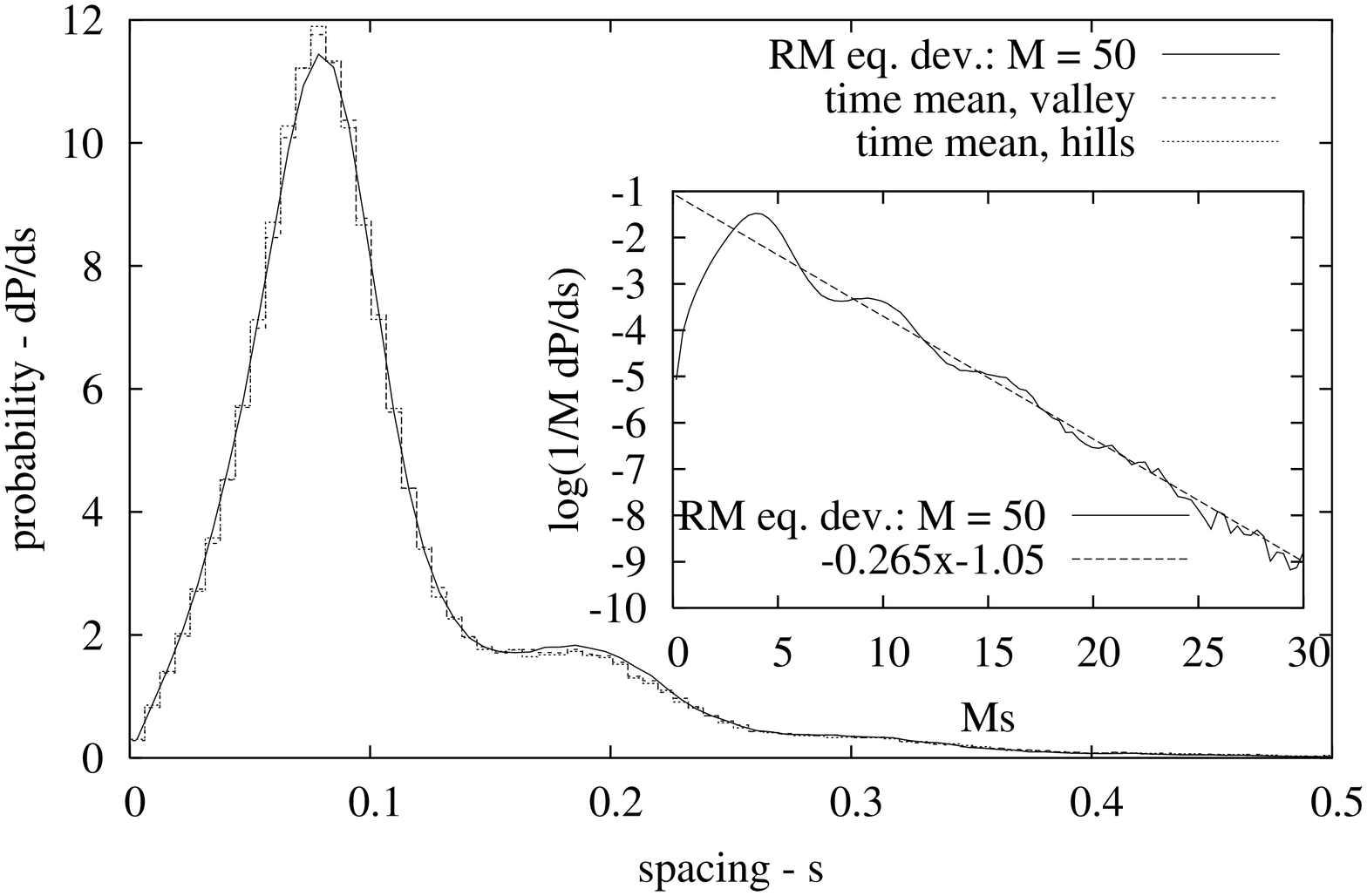}
    \includegraphics[angle=-90, width=7cm]{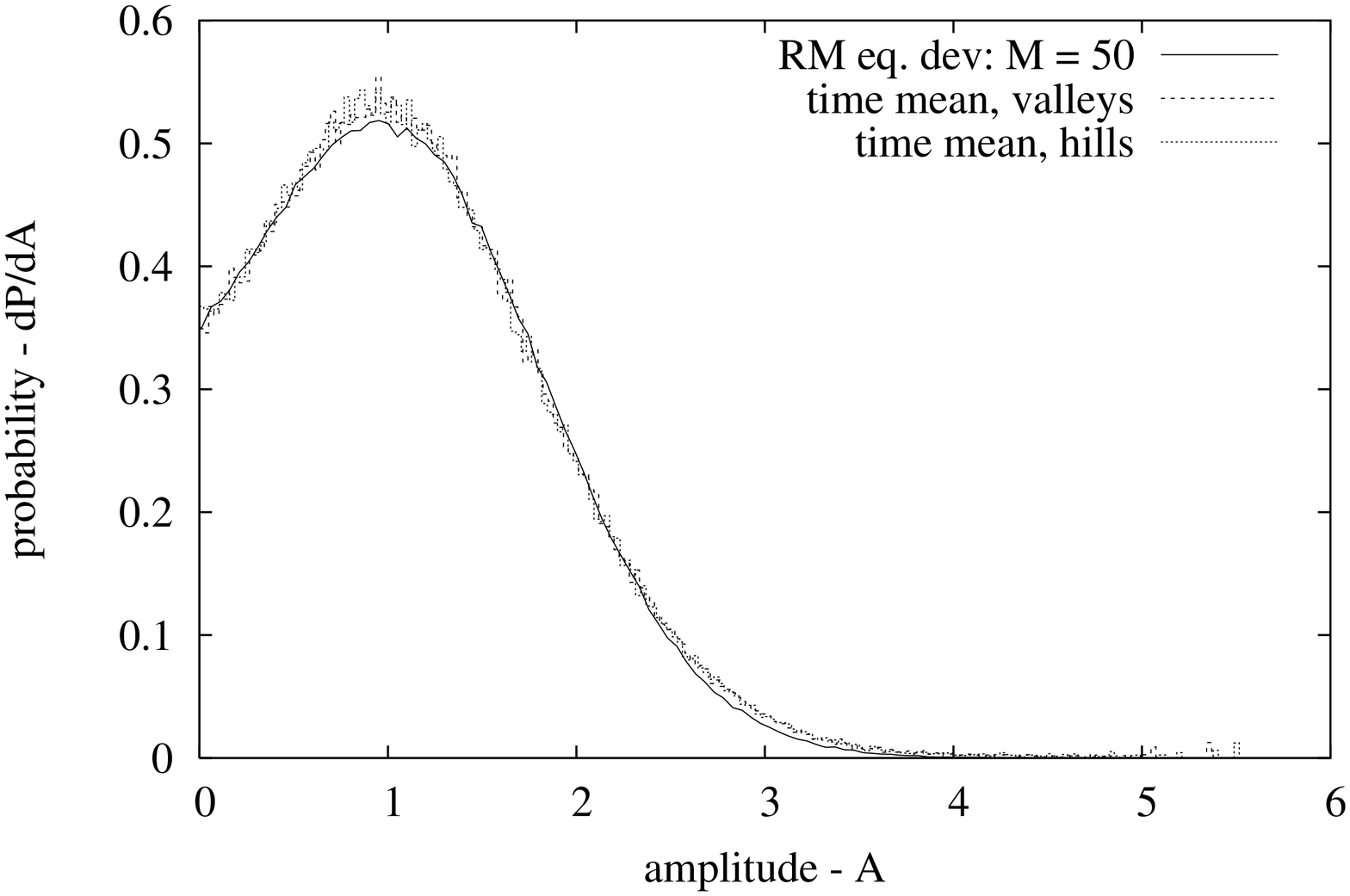}
    (c)\includegraphics[angle=-90, width=7cm]{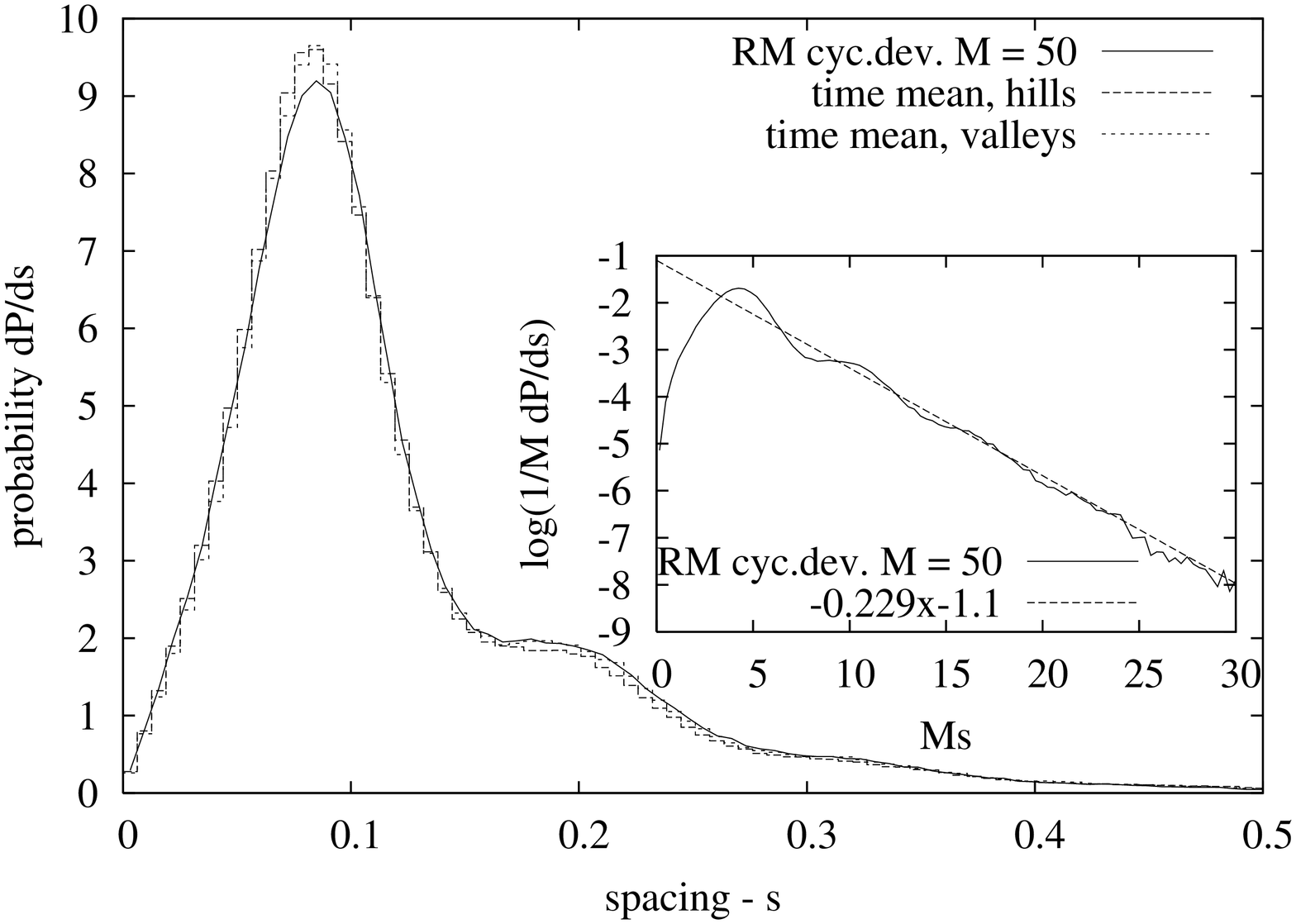}
    \includegraphics[angle=-90, width=7cm]{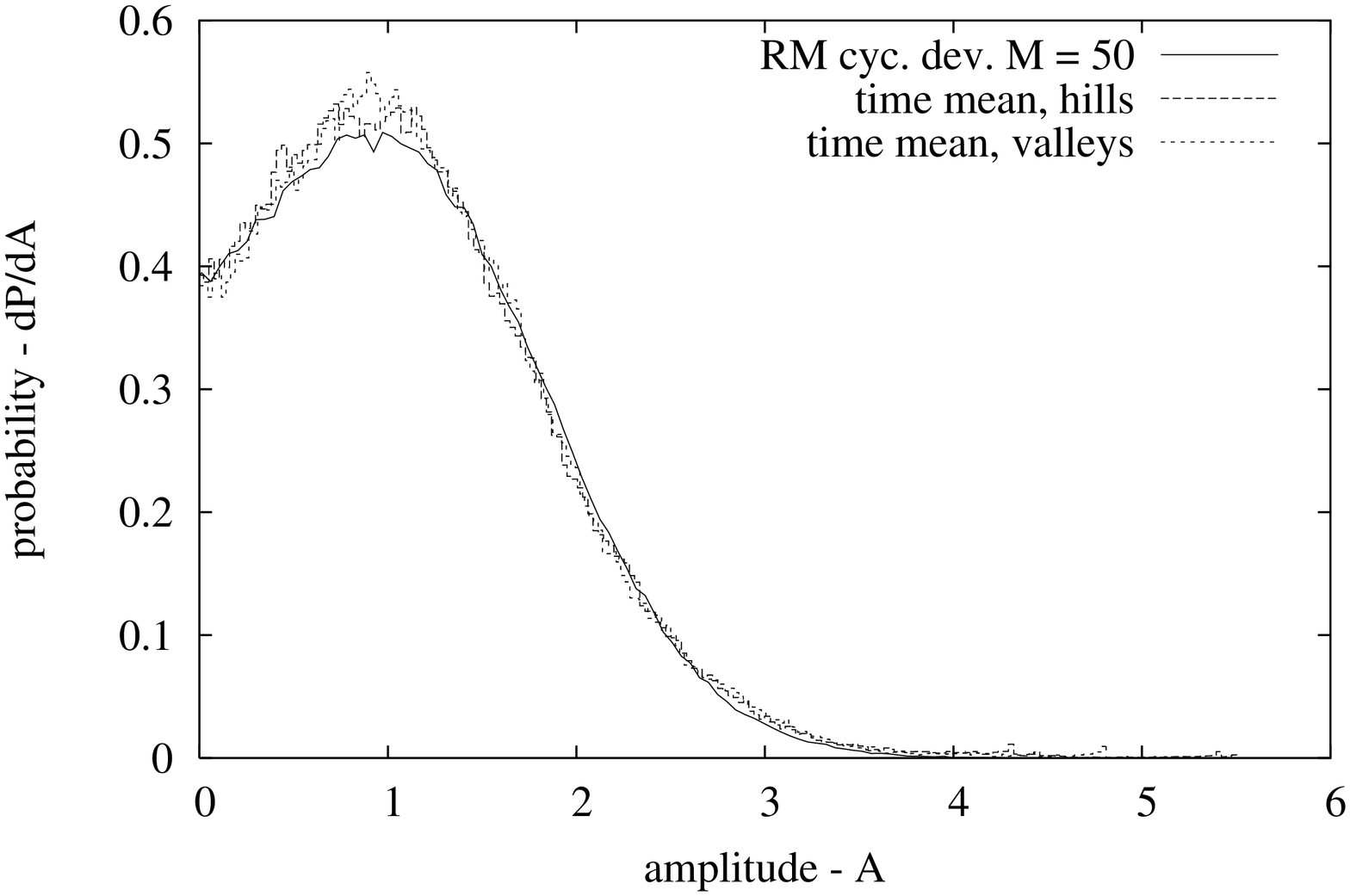}
  \end{center}
  \hfil (I) \hfil\hfil (II) \hfil 
  \caption{The average distribution of spacing $s$ (I) and amplitude $A$ (II) of WF along a random line in phase space: (a) discrete torus (see text for explanation), (b) continuous torus, (c) sphere. In the cases (b) and (c) the results of WFL simulations are compared with results obtained by simulations of the random model for the specific geometry. In all examples we consider the case with $M=50$ Fourier modes.}
\label{pic:str_WF}
\end{figure}

Results of our numerical studies of structure statistics $d{\cal P}/ds$, $d{\cal P}/dA$ and $d{\cal P}/dsdA$, are shown on the figure \ref{pic:str_WF}. Numerical experiments are done so that they correspond to WF random model with $M=50$. We can see a good agreement between the random model and numerical measurements on real dynamical systems. The latter are performed by time averaging over the evolution of an arbitrary initial state. We should stress that the numerical results are identical if we instead consider Wigner function of a random (ergodic) state. With this results we have again shown, indirectly, that the assumptions in the derivation of the random model are correct. In the case of a spherical phase space there is a small deviation of the spacing distribution $d{\cal P}/ds$ around the peak of distribution due to imperfect approximation of the mode variances with the asymptotic semicircle formula (\ref{eq:cycle_dev}).\par
The spacing distribution between neighboring zeros $d{\cal P}/ds$ of WFL (shown in figures \ref{pic:str_WF}I) gives some information on typical intersections of the nodal cells, whereas amplitude distribution $d{\cal P}/dA$  (shown in figures \ref{pic:str_WF}II) gives information on the distribution of the heights in such intersected filaments. However, we note an interesting observation, namely that the statistics of the structure of positive nodal cells ($W - \mean{W} > 0$) are identical to the statistics of negative nodal cells ($W-\mean{W} < 0$) in the asymptotic regime $N\to\infty$. This is consistent with an asymptotic symmetric Gaussian distribution of Wigner function values (\ref{eq:Gaussian}). The spacing distribution exhibits periodically spaced peaks, with period $\pi/M$, and exponential tail with non-universal exponent for large $s$ \cite{kuznetsov}.

\begin{figure}[!htb]
  \begin{center}
    \includegraphics[width=7cm]{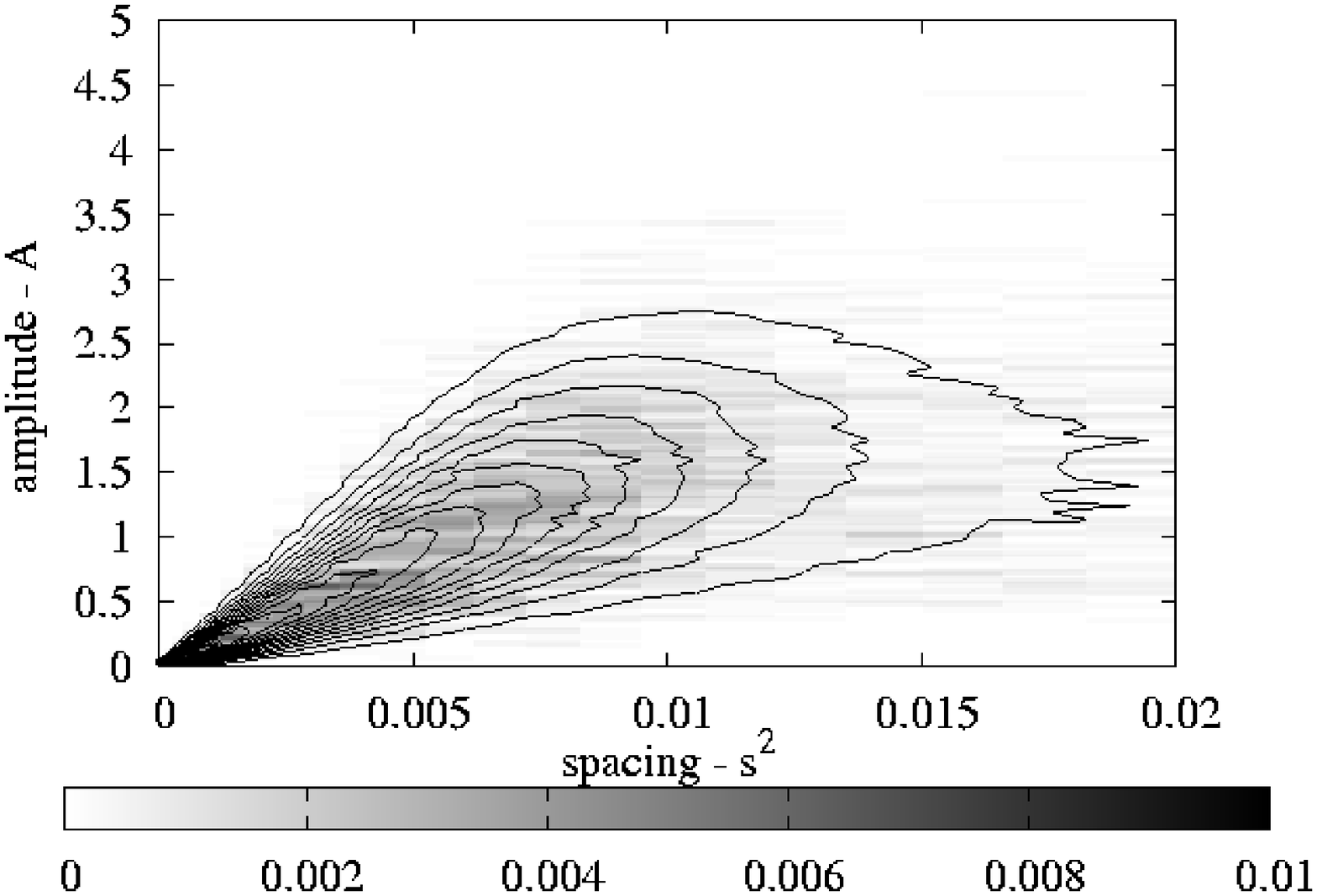}
    \includegraphics[width=7cm]{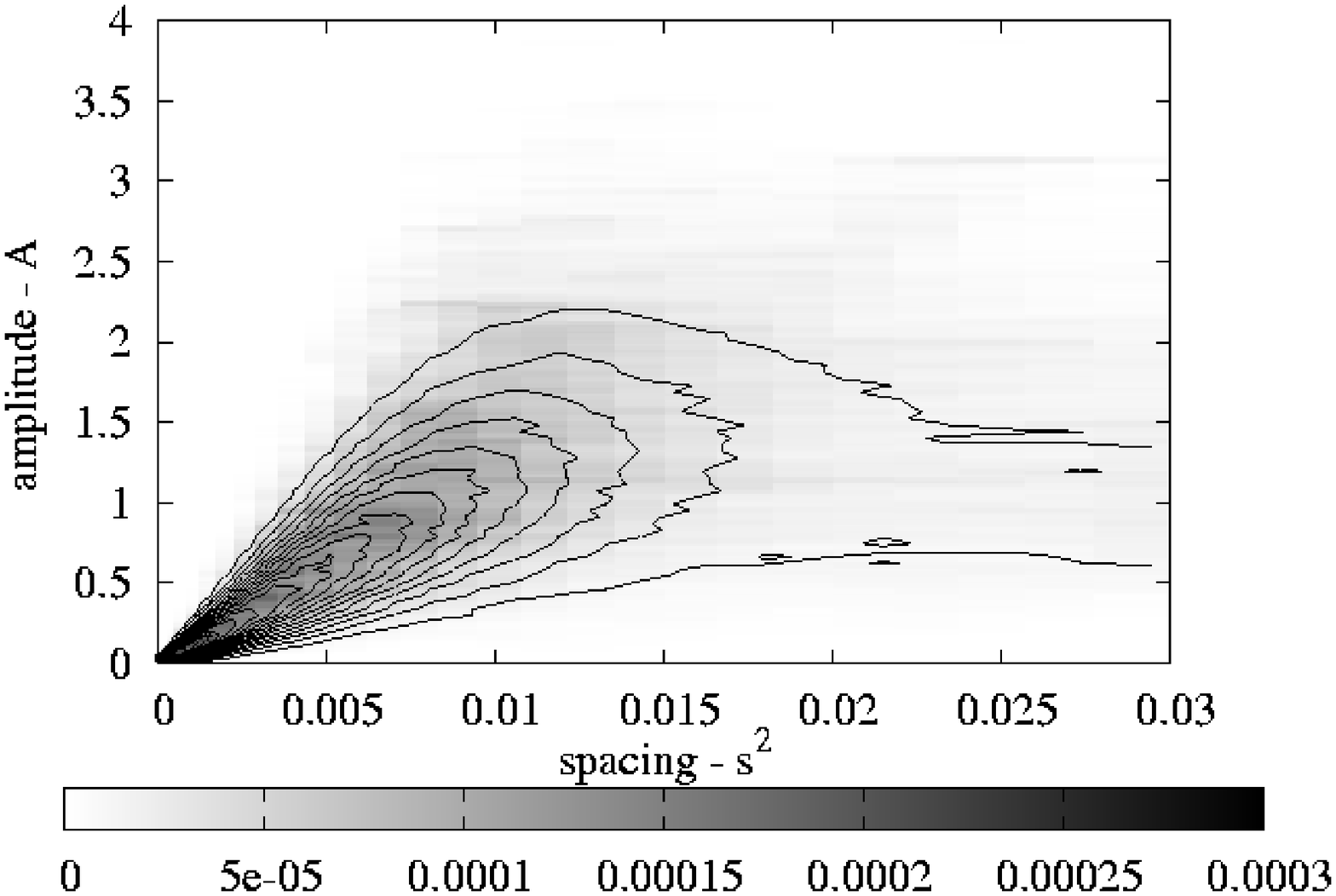}
  \end{center}
  \hfil (a) \hfil\hfil (b) \hfil 
  \caption{The average joint distribution of amplitude $A$ and spacing $s$ between two successive zeros of WFL: (a) continuous torus, (b) sphere. For more details see caption of figure \ref{pic:str_WF}.
The density plot with grayness scales (inicated in the horizotal bar) refers to the simulation of true WFL, whereas isodensity contours refer to random models as discussed in the text.
}
\label{pic:str_WF_sA}
\end{figure}

As the physical domain of Wigner function on a torus is in fact a discrete mesh of $N\times N$ points, we also consider statistics of a proper discrete WFL, namely ${\widetilde w}(m) = {\widetilde W}(2\pi m/N)$, and the discrete spacing distribution $P_s$ which in fact measure the distribution of clusters of points where ${\widetilde w}(m)$ has constant sign, and the corresponding amplitude distributions. These are shown in figure \ref{pic:str_WF}a. Interestingly, we find strong numerical evidence of the following exponential distribution 
\begin{equation}
P_s = 2^{-s}.
\end{equation}
On the other hand, this is in fact just the distribution of the lengths of clusters of repeated uncorrelated binary events. This is consistent with the statement that the values of WF of a random state on a physical quantum mesh of $N\times N$ phase space point are indeed uncorrelated as suggested already by the auto-correlation function.
We have also investigated the joint spacing-amplitude distributions $d{\cal P}(s,A)/dsdA$, shown in figure \ref{pic:str_WF_sA}. Again the results inferred from the simulations of the random model match the simulations of WFL. We found that distributions $d{\cal P}/dsdA$ have a simple shell-shaped form, indicating strong correlations between the spacing and the amplitude of the WF intersections. In fact one can give an upper bound on the maximally allowed amplitude $A$ at a given small spacing $s$, namely this can be estimated from the second derivative (curvature) of the random model ansatz (\ref{eq:model}).
\section{Summary and discussion}
In this paper we have proposed to study statistical properties of Wigner functions of random pure quantum states, either eigenstates or time-dependent states of classically chaotic systems. We concentrated on the properties of quantum maps with 2d classical phase space. We have shown, using simple general arguments, that the Wigner function value distribution of a random state should tend to a Gaussian, in the semi-classical limit, which is centered around zero value, so the probabilities to have a negative or positive value become equal in the limit. In other words, the standard deviation divided by the mean value of the Wigner function diverges as $\sigma/ \mean{W} \sim \sqrt{N}$ in the semi-classical limit $N\to\infty$. In addition, we have analyzed the structure and phase-space correlations of Wigner functions of random states. In particular, we have shown that the auto-correlation of the Wigner function becomes a delta function in the semi-classical limit, and that the nodal cells have typical structures on (sub-Planckian) scale $\delta q,\delta p \sim \hbar$.

We believe that our results may shed some new light onto the related studies of decoherence \cite{zurek,srednicki}. In particular, one now expect certain properties of the Wigner function of random states to be manifestly {\em non-classical}. For example, the fidelity of two initially equivalent states which undergo two slightly different time-evolutions 
has been found to behave non-classically for times larger than the $-\log\hbar$ Ehrenfest time, i.e. when the states become effectively random, whereas classical behavior of fidelity has been recovered for shorter times \cite{PZ02}. We stress that we have in this study only considered non-autonomous (time-dependent, e.g. kicked) quantum systems, where Wigner functions of ergodic states are expected to occupy the entire classical phase space. On the other hand, one may ask similar questions about the statistics of Wigner functions on, or close to, energy surfaces of autonomous Hamiltonian systems. This is a subject of a forthcoming publication \cite{PB02}.
%
\section*{Acknowledgments}
Useful discussions with A. B\" acker, G. Veble and M. \v Znidari\v c, as well as the financial support by the Ministry of Education, Science and Sport of Slovenia are gratefully acknowledged.

\end{document}